\documentclass[12pt]{article}

\newcommand{\LI}{\hbox to\hsize}
\newcommand{\LLI}[1]{\LI{#1\hss}}

 1
 1
 1

\newcommand{\PM}[1]%
{\mbox{$m_{\rm #1}$}} 
\newcommand{\BEQ}{\begin{equation}}
\newcommand{\EEQ}{\end{equation}}
\newcommand{\abs}[1]{\mbox{$\left| #1 \right|$}} 

\newcommand{\lappr}{\mbox{$\stackrel{<}{\sim}$}} 
\newcommand{\incircle}[1]{\mbox{{\hbox{$\bigcirc$}\kern-0.7em
\lower0.05ex\hbox{\mbox{{\scriptsize\rm #1}}}}}}
\newcommand{\eq}[1]{eq.~(\ref{#1})}
\newcommand{\ETC}{\mbox{\em etc.\/ }}

\newcommand{\IE}{\mbox{\em i.e. \/}}
\newcommand{\ETAL}{\mbox{\em et. al.\/ }}
\newcommand{\EG}{\mbox{\em e.g.\/ }}
\newcommand{\tipac}[2]{\begin{flushright}JHU--TIPAC #1\\
#2 
\end{flushright}}
\newcommand{\titleby}[3]{\vskip25truemm \begin{center}
{\large\bf #1}\\[4truemm] #2\\[2mm]
Department of Physics and Astronomy\\
The Johns Hopkins University\\
Baltimore, MD 21218\footnote{E-mail: #3}\\  \end{center}}
\renewcommand{\abstract}[1]{\vskip20truemm \begin{quote} #1
 \end{quote} \vskip5mm}
\newcommand{\PACS}[1]{\begin{flushleft} PACS: #1 \end{flushleft} \vskip5mm}

\begin{document}
\input epsf
\tipac{96001}{February 1996}
\titleby{Neutrino Moments and the Magnetic Primakoff Effect}{G.~Domokos\\
and\\
S. Kovesi--Domokos}{SKD@HAAR.PHA.JHU.EDU}
\abstract{If  different species of neutrinos possess transition
magnetic moments, a conversion between species can occur in the 
Coulomb filed of a
nucleus. In the case of Dirac neutrinos this corresponds to an active to
sterile conversion, whereas in the case of Majorana neutrinos, the
conversion takes place between active species. The conversion cross
sections grow with the energy of the incident neutrino. The formalism
is also  applied to  a new type of experiment designed to test 
the existence of the ``KARMEN anomaly''.}\vskip3mm
The question whether neutrinos possess magnetic moments (both flavor diagonal 
and/or
transition moments) is of importance in exploring physics  beyond the
Standard Model. There exists a large number of theoretical models in which
neutrinos possess magnetic moments; also, a variety of upper limits has been
placed on the magnitude of moments neutrinos of various flavors can possess.
Typically, the upper bounds obtained from the various experiments and
observations give $\mu \lappr 10^{-11} \mu_{B}$, where $\mu_{B}$
stands for the Bohr magneton, with a varying degree of model dependence.
For a recent review see \EG \cite{kimpevsner}.

Here  we propose a novel method by means of which transition
moments can be measured. In essence, we propose to take advantage 
of the magnetic analogue of the Primakoff effect \cite{primakoff}: if a 
neutrino possesses a {\em transition} moment, the Coulomb field
of a nucleus will induce a helicity flipping transition dominantly between the
incoming neutrino and one for which the transition moment is largest. 
(In the case
of three flavors, the transition moment is a $3\times 3$ matrix in
flavor space. However, for the purposes of this work, we tentatively
assume that just one moment dominates the transition.) The  observed 
effect depends on whether neutrinos are Majorana or Dirac
particles. In the case of Majorana neutrinos, the incoming neutrino
flips to an active one of a different flavor. Due to the fact that 
neutrino beams consist mostly of muon neutrinos, the Coulomb field
will induce an excess of electron neutrinos and/or tau neutrinos. Those, 
in turn, 
will create electrons or taus in a neutrino detector. By contrast, if
neutrinos are Dirac particles, the magnetic transition in the Coulomb field
converts them into right handed, sterile neutrinos. Hence, there will be 
a depletion of the active muon neutrinos from the beam.

As another application of this effect, we examine 
the production of the particle conjectured by the  KARMEN collaboration.
An anomaly in the time distribution
of neutrinos measured at ISIS \cite{KARMEN} was interpreted as 
an ``$x$--particle'' of mass $M\approx 34$MeV. The dominant decay channel
in ref.~\cite{KARMEN} was conjectured to be 
\[  x \rightarrow \nu + \gamma 
\label{nustar}
\]

An independent experiment, performed at the Paul Scherrer Institute \cite{PSI}
failed to confirm the original KARMEN result.
Moreover, the original  interpretation is not without  problems of its own 
see \EG
the paper of Barger \ETAL \cite{barger}. In essence, the authors of 
ref.~\cite{barger} argue that if the $x$-particle exists, it can only
be a sterile ``neutrino'' mixing with an active one.  For this reason, it
is of importance to test the existence of the ``$x$--particle'' in a
different type of experiment. If indeed $x$ is produced by muon neutrinos in
the Coulomb field of a nucleus, the photons resulting from its decay may
be observed downstream.

We give
formulae for the  differential and total cross sections. The calculation
is carried out both for a spin 1/2 and spin 3/2 particle in the
final state. 
There are various reasons why the possibility of
a spin 1/2 $\rightarrow$ spin 3/2 transition may be of interest. 
We mention a few.

In a variety of theories there may exist
neutral particles of spin 3/2 which mix with particles of spin 1/2; 
 arguments why a  spin 3/2 particle should be much heavier
than a neutrino or the particle allegedly observed by the KARMEN
collaboration are mostly based on the fact that, apparently,
the fact that transitions to spin 3/2 leptons have not
been observed. In addition, the idea that quarks and
leptons possess a substructure keeps recurring in the literature.
If a substructure indeed existed, one should consider the possibility
of excited quarks and leptons. If the known spectrum of hadrons
can serve as a guide, one expects the lowest lying fermionic excited state
to have spin 3/2.

Admittedly, neither one of these motivations is 
currently a very strong one; nevertheless, one should keep an open mind
about the possibilities.
Moreover, the measured partial width of the Z into
invisible channels places very severe limitations onto the
couplings (mixing angles, respectively) of such particles.

If the transition is
between states of  spin 1/2 and spin 3/2, existing limits on transition moments
are somewhat weakened and, hence, some of the arguments 
in ref.~\cite{barger} are
weakened too, since a  barrier 
penetration factor suppresses the transition rates. For most of the
processes in question the suppression is not sufficiently dramatic
in order to warrant attention. The only exception is the decay,
$ \pi \rightarrow \mu + x.$
In that case, due to the very small Q value of the decay,
the barrier penetration factor is about $10^{-3}$ and correspondingly,
the branching ratio is suppressed by that factor.

We recall that in the case of the original Primakoff effect (\IE
photoproduction of neutral pions in the Coulomb field of a nucleus),
the effective interaction is related to the anomaly of the 
neutral isovector axial current, {\em viz}
\BEQ
{\cal L}_{eff} = \frac{\alpha}{4 \pi f_{\pi}}\pi^{0}f_{\mu \nu}F^{\mu \nu},
\label{primakoff}
\EEQ
where $f_{\mu \nu}$ and $F_{\mu \nu}$ stand for the field tensor describing
the incident photon and the field tensor of the (screened) Coulomb
field, respectively.

In a similar fashion, the effective Lagrangian describing the 
transition between two neutrino species, say, $\nu_{1}$ and 
$\nu_{2}$ is proportional to $F_{\mu \nu}$. 
We have the following expressions.
 
\LLI{Transition between spin 1/2 neutrinos:}

\BEQ
{\cal L}^{s=1/2}_{eff} = \frac{1}{2} \mu_{B} \overline{\nu_{2}}
\left( \kappa_{n} + \kappa_{a} \gamma_{5} \right)
\sigma_{\rho \sigma} \nu_{1} F^{\rho \sigma} + h.c.
\label{spin1/2} 
\EEQ
 
\LLI{Transition between spin 1/2 and spin 3/2 neutrinos:}

\BEQ
{\cal L}^{s=3/2}_{eff} = 
\frac{1}{2Mi} \mu_{B} \overline{\nu_{2}^{\lambda}}
\left( \kappa^{'}_{n} + \kappa^{'}_{a} \gamma_{5} \right)
\sigma_{\rho \sigma} \nu_{1} \partial_{\lambda}F^{\rho \sigma}  + h.c.
\label{spin3/2} 
\EEQ

In the equations above, the quantities $\kappa$ stand for the effective
coupling strengths measured in units of a Bohr magneton,
$\mu_{B} = e/2m_{e}$ and M stands for the mass of the neutrino of
spin 3/2.
The subscripts $ n, a$ refer to normal and
anomalous transitions, respectively. Finally, $\nu^{\lambda}$ stands for
a Rarita--Schwinger spinor describing a neutrino of spin 3/2.

It is to be noted, however that in the experiments discussed here, one
cannot distinguish between the cases of normal and anomalous
transitions and, apart from the interpretation of the experiments 
outlined above, between Majorana and Dirac neutrinos either. 
As an illustration
consider, \EG the inverse lifetime of a heavy, spin 1/2 neutrino 
decaying into
a light $\nu$ and a $\gamma$. Apart from a common
phase space factor, it is given by the expression:
\[
\Gamma (\nu_{2} \rightarrow \nu_{1} + \gamma)
\propto \left( \abs{C_{n}}^{2} + \abs{C_{a}}^{2}\right),
\]
where
\[ \abs{C_{n}}^{2} = \abs{\kappa_{n}}^{2}, \quad
\abs{C_{a}}^{2} = \abs{\kappa_{a}}^{2}, \]
for Dirac neutrinos, whereas
\[ \abs{C_{n}}^{2}= 4 \left( Im \kappa_{n}\right)^{2}, \quad
\abs{C_{a}}^{2} = 4 \left( Re \kappa_{a}\right)^{2}\]
for Majorana neutrinos. 
The situation is similar in the case of
{\em unpolarized} cross sections and for transitions involving
excited neutrinos of spin 3/2. 

The electromagnetic interaction conserves parity, hence, either
$C_{n}$ or $C_{a}$ vanishes, depending on the relative parities
of the neutrinos in the initial and final states. However,
the amplitudes of normal and anomalous transitions differ in
terms which are  of the order of magnitude of the neutrino
masses involved. (For {\em small} masses a neutrino is {\em almost}
an eigenspinor of $\gamma_{5}$.)
Henceforth, we arbitrarily set $\kappa_{a}=\kappa^{'}_{a}=0$
for the rest of this paper; the effective couplings will be denoted by
$\kappa$ and $(\kappa^{'}$, respectively and we express our results in terms
of a single  effective low energy coupling constant. Moreover, we can
disregard differences between Dirac and Majorana neutrinos.

The calculation of the differential cross section of the process
\[ \nu_{1} \rightarrow \nu_{2} \]
in a screened Coulomb field and using the effective interaction
\eq{spin1/2} or \eq{spin3/2} respectively, is an elementary exercise.
The only expression not given in elementary texts is the helicity sum
over Rarita--Schwinger spinors, as given \EG in ref.~\cite{scadron}.
We do not need the full expression, only its contraction with
$q_{\alpha}$, the momentum of the virtual photon. 
We define:
\BEQ
P_{\alpha \beta}\left( p \right) =
\sum_{\Lambda = - 3/2}^{3/2} \nu_{\alpha}^{\Lambda}\left( p \right)
\overline{\nu_{\beta}^{\Lambda}\left( p \right)},
\EEQ
where $\Lambda$ is the helicity of the spin 3/2 neutrino.

One finds for instance, for $p_{0} > 0$:
\BEQ
b = {P}_{\alpha \beta} q^{\alpha} q^{\beta} = \frac{2}{3}
\left[ \frac{\left( p\cdot q\right)^{2}}{M^{2}} - q^{2}\right]
\left( \not{p} + M \right).
\label{barrier}
\EEQ
(One easily verifies that in a decay, \EG $\nu_{2} \rightarrow 
\nu_{1} + \gamma$ the coefficient of $(\not{p} + M)$ in the
expression of  $b$ is just
 a conventional barrier penetration
factor.)

We  quote the differential and total cross sections for spin 1/2 and 3/2
neutrinos in the final state. The formulae are written down in the high
energy limit: the beam energy is much larger than the mass of the neutrino
either in the initial or in the final states. The differential cross sections
are given in the laboratory frame.
We have:
\LLI{For $s=1/2$ to $s=1/2$ transition:}
\BEQ \frac{d\sigma^{1/2}}{d\Omega} \sim 8\pi Z^{2}\alpha \kappa^{2}
\mu_{B}^{2} \frac{\sin^{2}\theta /2}{\left( \sin^{2}\theta /2 
+ \frac{\mu^{2}}{4E^{2}}\right)^{2}}
\label{diff1/2}
\EEQ
\BEQ \sigma_{t}^{1/2} \sim 8\pi^{2} Z^{2}\alpha \kappa^{2} \mu_{B}^{2}
\left( -1 + 2\ln \frac{2E}{\mu}\right)
\label{tot1/2}
\EEQ
\LLI{For $s=1/2$ to $s=3/2$ transition:}
\BEQ
\frac{d\sigma^{3/2}}{d\Omega} \sim \frac{16\pi}{3} Z^{2}\alpha \kappa^{2}
\mu_{B}^{2} \left(\frac{E}{M}\right)^{4}\sin^{2}\theta /2
\label{diff3/2}
\EEQ
\BEQ \sigma_{t}^{3/2}\sim \frac{32 \pi^{2}}{3} Z^{2}\alpha \kappa^{2}
\mu_{B}^{2} \left(\frac{E}{M}\right)^{4}
\label{tot3/2}
\EEQ

In these equations, $\mu$ stands for the inverse of the
screening radius. Using the Thomas--Fermi model of atoms,
it is given by the expression, $\mu = a_{0}^{-1}Z^{-1/3}$, where $a_{0}$
is the Bohr radius. (In energy units $ \mu \approx 3.65 \times 10^{- 3}
Z^{-1/3}$MeV.)
We notice that in the expressions involving a
spin 3/2 final state, the screening radius does not appear in the 
asymptotic expressions, \eq{diff3/2} and \eq{tot3/2}. This is due to
the presence of the barrier penetration factor appearing in  $b$.

For angles down to $\theta \simeq \mu /E$, the angular
distribution given by \eq{diff1/2} behaves essentially as
$ 1/\sin^{2}\theta/2$. The angular distribution
given by \eq{diff3/2} is quite flat, again, due to the presence of the
barrier penetration factor.

There is a radical difference between the energy dependence of the
total cross sections. Both the spin 1/2 and spin 3/2 cross sections grow
with the beam energy $E$, but the spin 3/2 cross section grows much more
rapidly. For this reason, it is important to estimate the critical energy
at which \eq{tot3/2} violates the unitarity bound. 

The absolute value of the scattering amplitude is given by:
\[ \left| f^{3/2}(\theta) \right| = 
\left(\frac{d \sigma^{3/2}}{d \Omega}\right)^{1/2} \]
From here we find the estimate for the S-wave phase shift:
\[ \left| \sin \delta_{0} \right| \sim Z
\sqrt{\frac{8\pi}{3}\alpha}\frac{2^{3/2}}{3} \mu_{B}\kappa M
\left( \frac{E}{M} \right)^{3} \]
At the critical energy, $E_{c}$, $\left| \sin \delta_{0}\right| \approx 1$.
Depending on the estimates on $\kappa$ and $M$ (see \cite{kimpevsner}),
one gets various critical energies. Using conservative
estimates taken from that reference, one gets 
$E_{c}\approx 10^{6}$GeV or so. This value of $E_{c}$ is comfortably
high for the purposes of any terrestrial beam. However, neutrinos
emerging from active galactic nuclei may have comparable
energies, {\em cf.} \EG\cite{hawaii}. For that reason, more
careful estimates are needed (using $K$--matrix unitarization \ETC)
for the highest energies.
 
From the experimental point of view, if the particle observed by the 
KARMEN collaboration exists (and its dominant decay mode is
what has been conjectured by the collaboration), its spin can be 
determined from the energy
dependence of its production  cross section {\em via} the magnetic
Primakoff effect. In general, the energy and Z dependence of the conversion
probability provide  important clues  in determining whether a conversion 
occurs due to
oscillations or due to the magnetic Primakoff effect. In fact, the
oscillation length {\em increases} linearly with the neutrino energy
(except perhaps in the neighborhood of a MSW resonance), whereas
the comparable quantity, the mean free path of the magnetic
Primakoff effect {\em always decreases with the neutrino energy\/:}
it is, roughly, proportional to $1/\ln E$ or $1/E^{4}$depending on the
spins of the particles involved, {\em cf\/.} the expressions of the
cross sections.
In Fig.~1  we displayed the total 
cross section as a function of 
the energy. We remark that for moderate energies, 
($E\approx 15$GeV) the spin 1/2 to spin 1/2 transition has a 
cross section comparable to a weak cross section within the
framework of the standard model (see \cite{quigg}) for 
$ \kappa \approx 2 \times 10^{-8}/Z$. 
 \begin{figure}  
 \begin{center} 
 \epsfbox[72 214 540 578]{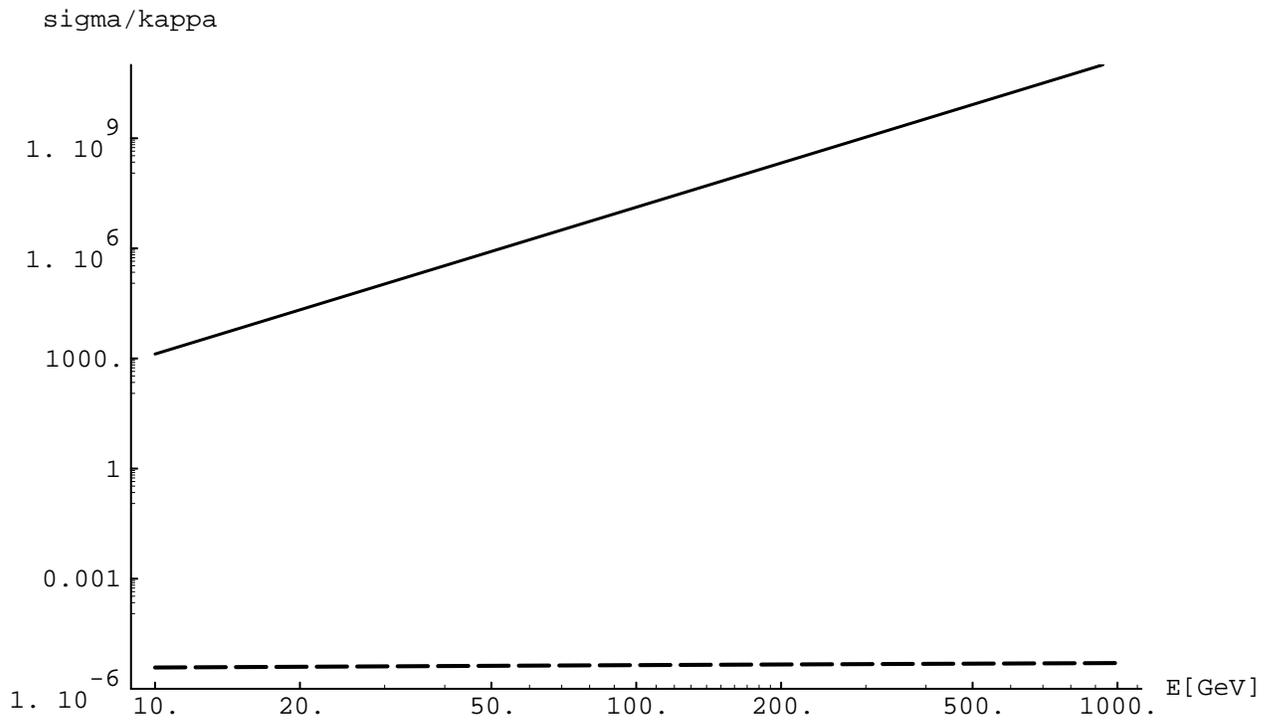}
 \end{center}
 \caption{\small  Plot of the cross sections of the magnetic
 Primakof effect as a function of the energy. The cross sections are
 plotted for Z=26. Dashed line: spin 1/2 final state. Full line:
 spin 3/2 final state. 
 The symbols labeling the axes mean the following: 
 \mbox{sigma=$\sigma/10^{-10}{\rm mb}$}, \mbox{kappa=$10^{11}\kappa$}
 The cross section of the spin 1/2 to spin 3/2 transition was 
 plotted for
 M=34MeV} 
 \end{figure}

We notice that the transition to a spin 3/2 final state has a
much larger cross section than the one to spin 1/2. The fact that, in all
probability, no such transition has been detected so far, puts very stringent
upper limits on the transition moments. (We have not yet analyzed existing
data from this point of view.)

A further severe limitation on the existence (more precisely, the couplings) of
a light spin 3/2 particle arises from the measured invisible
decay width of the Z.
The decay width of the Z into a pair of spin 1/2 particles with
$ M\ll M_{\rm Z}$ is given by the expression:
\BEQ
\Gamma ({\rm Z}\rightarrow \nu + \overline{\nu}) \approx
\frac{G_{\mu} M_{\rm Z}^{3}}{12 \sqrt{2}}\left(g_{L}^{2}
+ g_{R}^{2}\right).
\label{Zdecay}
\EEQ
Here $G_{\mu}$ is the muon decay constant, and $g_{L}$ and $g_{R}$
are the {\em effective} left and right handed couplings of the Z to the
hypothetical extra neutrino, respectively.
Assuming that the error on the invisible width \cite{particledata} 
of the Z leaves room for an extra neutrino, one gets the upper limit,
\[ \sqrt{g_{L}^{2} + g_{R}^{2}}\quad \lappr \quad 5 \times 10^{-3}. \]
However, if the extra neutrinos had spin 3/2, the expression
\ref{Zdecay} has to be multiplied by a factor, 
\[ \frac{1}{36} \left( \frac{M_{\rm Z}}{M}\right)^{8}. \]
(This factor arises from \eq{barrier} in the limit 
$M_{\rm Z}\gg M$.)

For all practical purposes therefore, the coupling of the Z to a light,
spin 3/2 particle is zero. Of course, in the absence of a compelling
theory, there is no {\em a priori} reason to exclude the existence
of such a particle. However, if it exists, it is extremely unlikely that it is
coupled to any of the known particles. 

After the calculation reported here was finished, we became aware of
the work described in ref.~\cite{vannucci}. The authors of that 
work developed ideas similar to the ones described here, but in 
a rather different context.

We thank F.~Halzen for drawing our attention to ref.~\cite{vannucci}
and for encouraging us to publish this result. We also thank
Leon Madansky and Barry Blumenfeld for useful discussions.

\end{document}